\documentclass[11pt,a4paper]{article}
\usepackage{jinstpub}

\usepackage{amsmath}
\usepackage[numbers,sort&compress]{natbib}
\usepackage{graphicx}
\usepackage{color}
\usepackage{lineno}
\usepackage{booktabs}
\usepackage{multirow}
\usepackage[euler]{textgreek}
\usepackage{float}
\usepackage{orcidlink}

\usepackage{savesym}
\savesymbol{tablenum}
\usepackage{siunitx}
\usepackage{hyperref}
\restoresymbol{SIX}{tablenum}
\usepackage{tabularx}
\usepackage{xcolor}
\let\svthefootnote\thefootnote
\newcommand\blankfootnote[1]{%
  \let\thefootnote\relax\footnote{#1}%
  \addtocounter{footnote}{-1}\let\thefootnote\svthefootnote
}

\title{Automated analysis of the visual properties of superconducting detectors}

\author[a,b,c,1]{K.~R.~Ferguson,%
 \orcidlink{0000-0002-4928-8813}%
 \note{Corresponding author.}}
\author[b,d,e]{A.~N.~Bender,%
 \orcidlink{0000-0001-5868-0748}}
\author[c]{N.~Whitehorn,%
 \orcidlink{0000-0002-3157-0407}}
\author[f]{P.~S.~Barry,}
\author[b]{T.~W.~Cecil,%
 \orcidlink{0000-0002-7019-5056}}
\author[d,e]{K.~R.~Dibert,}
\author[d,e]{and E.~S.~Martsen}

\affiliation[a]{Department of Physics and Astronomy, University of California Los Angeles,\\
475 Portola Plaza, Los Angeles, CA 90095, USA}
\affiliation[b]{High-Energy Physics Division, Argonne National Laboratory,\\
9700 South Cass Avenue, Lemont, IL 60439, USA}
\affiliation[c]{Department of Physics and Astronomy, Michigan State University,\\
567 Wilson Road, East Lansing, MI 48824, USA}
\affiliation[d]{Department of Astronomy and Astrophysics, University of Chicago,\\
5640 South Ellis Avenue, Chicago, IL 60637, USA}
\affiliation[e]{Kavli Institute for Cosmological Physics, University of Chicago,\\
5640 South Ellis Avenue, Chicago, IL 60637, USA}
\affiliation[f]{School of Physics and Astronomy, Cardiff University,\\
Cardiff CF24 3YB, United Kingdom}

\emailAdd{kferguson@physics.ucla.edu}

\abstract{The testing and quality assurance of cryogenic superconducting detectors is a time- and labor-intensive process. As experiments deploy increasingly larger arrays of detectors, new methods are needed for performing this testing quickly. Here, we propose a process for flagging under-performing detector wafers before they are ever tested cryogenically. Detectors are imaged under an optical microscope, and computer vision techniques are used to analyze the images, searching for visual defects and other predictors of poor performance. Pipeline performance is verified via a suite of images with simulated defects, yielding a detection accuracy of $98.6\%$. Lastly, results from running the pipeline on prototype microwave kinetic inductance detectors from the planned SPT-$3$G$+$ experiment are presented.
}

\keywords{computer vision, cryogenic detectors, MKIDs}

\begin{document}

\maketitle
%\linenumbers

%Cosmology aims to answer some of the largest open questions in physics today; questions relating to the nature of dark matter and dark energy, the details of neutrinos, and even inflationary and GUT-scale physics~\citep{arbey21, hoekstra08, abazajian15b, vazquez18, mohanty08}. The probes that cosmologists use to study these problems are myriad, and include galaxy clusters, gamma ray bursts, and supernovae~\citep{allen11, moresco22, goobar11}. Of particular interest as a cosmological probe is the cosmic microwave background (CMB), the relic radiation from the Big Bang~\citep{abazajian15a, abazajian16}. As the oldest light in the universe, the CMB is the best tool available to study the universe in its infancy.

%The current state-of-the-art in the CMB field is to use superconducting detectors to perform measurements of the microwave sky. Most modern experiments use transition-edge sensor (TES) bolometers, though upcoming experiments such as SPT-$3$G$+$ are planning to use microwave kinetic inductance detectors (MKIDs). Kinetic inductance is additional inductance gained by a superconducting material when its Cooper pairs are broken. MKIDs make use of this by coupling a superconducting inductor with inductance $L$ to a capacitor with capacitance $C$, creating a circuit with a specific resonant frequency $f_{0} = \left( 2\pi \sqrt{LC} \right)^{-1}$. As incident photons impart power and break Cooper pairs, the inductance of the superconductor changes, inducing a shift in the resonant frequency that can be read out to determine the sky signal~\citep{day03}.

Cryogenic superconducting detectors are used in a wide array of scientific and technological applications, including quantum computing, particle physics, biomedicine, and astronomy \citep{nam04, pretzl20, previtali06, mokbel24, brink12, rando04}. Though different types of superconductors may be chosen for different applications, these detectors are broadly applicable due to their low noise. Current state-of-the-art detectors are generally dominated by the intrinsic uncertainty in the arrival rate of incident photon stream. In some applications, such as observation of the cosmic microwave background, increasingly larger numbers of detectors and arrays are being developed and deployed to increase overall sensitivity and facilitate higher-precision measurements.

%Because this noise decreases with the number of photons collected, larger and larger arrays of these detectors are developed and deployed to collect more photons at a faster rate, decreasing the overall noise and facilitating higher-precision measurements.

%For example, the first camera on the South Pole Telescope (SPT), SPT-SZ, comprised only $960$ TES bolometers~\citep{chang09, shirokoff09}; the current camera, SPT-$3$G, comprises approximately $16{,}000$ detectors~\citep{sobrin22}, while SPT-$3$G$+$ will use approximately $34{,}100$ detectors~\citep{anderson22}. Future experiments, such as CMB-S$4$, will continue to see this number increase to $\mathcal{O} \left( 500{,}000 \right)$ detectors~\citep{barron22}.

The fabrication and testing of increasingly larger arrays of superconducting detectors poses a significant undertaking. Testing throughput is often limited by the cooldown times of cryogenic hardware. Because the detectors must be cooled to extremely low temperatures to function, testing even a small number of detectors can often be a long (i.e., multi-week) process. It is therefore desirable to flag detectors that are likely to not perform well \textit{before} this testing is done in order to not waste time testing nonfunctional detectors.

One potential indicator of poor performance is the presence of visual defects on the detector. Much recent work has been performed to advance the field of automated anomaly detection, usually focusing on industrial manufacturing (for a comprehensive review, we refer the interested reader to~\citet{liu24}). Current state-of-the-art methods employ deep learning techniques to search for defects in images of manufactured objects. While some supervised models exist \citep{qiu19, baitieva24}, most work focuses on unsupervised learning due to the relative scarcity of anomalous objects on industrial production lines and the relative difficulty of producing pixel-level anomaly masks to help train the model. In particular, methods such as \citet{lee22} and \citet{batzner24} demonstrate promising performance on the MVTec Anomaly Detection dataset \citep{bergmann21}, a standard dataset used for benchmarking.

In this paper, we propose a novel method of using more traditional computer vision techniques to flag likely under-performing detector chips at room temperature --- before they are ever packaged and installed in a cryostat. As larger detector arrays are produced and tested for deployment-readiness, this process will reduce time spent testing wafers that have a low probability of being used in an experiment. To validate our framework, we search for defects that would cause an inoperable detector or a shift in resonant frequency in prototype microwave kinetic inductance detectors (MKIDs) from the planned SPT-$3$G$+$ experiment. While the previously mentioned deep learning techniques are powerful, it can often be more straightforward and less computationally intensive to use more traditional computer vision techniques when searching for defects of known types on known geometries. In particular, many common techniques struggle with finding small defects in extremely large images, as is the problem at hand with SPT-$3$G$+$ detectors. While this shortcoming can be mitigated somewhat by splitting images into smaller tiles, one then loses global context that can also be important in determining detector behavior.

The paper is outlined as follows. In Section \ref{sec:hardware}, we briefly introduce SPT-$3$G$+$ and describe the detectors used to test this technique. In Section \ref{sec:imaging}, we describe the imaging and post-processing procedures. In Section \ref{sec:pipeline}, the analysis pipeline is described in-depth. Section \ref{sec:simulations} follows with a description of the simulation framework used to verify pipeline performance. Finally, Section \ref{sec:results} contains the results of running the pipeline on two test chips of SPT-$3$G$+$ MKIDs.

\section{SPT-3G+ MKIDs}
\label{sec:hardware}
SPT-$3$G$+$ is a camera currently being developed for deployment on the South Pole Telescope (SPT), which observes the microwave sky, including the cosmic microwave background (CMB). The camera is being designed to observe the secondary anisotropies of the CMB across three bands centered at $220$~GHz, $285$~GHz, and $345$~GHz using ${\sim} 34{,}000$ MKIDs with operating frequencies spanning from about $0.73$ to $1.33$~GHz~\citep{anderson22}. These detectors function on the principle that the kinetic inductance of a superconducting material changes when its Cooper pairs are broken by incident photons of sufficient energy. MKIDs take advantage of this phenomenon by coupling a superconducting inductor to a capacitor, creating a resonant $LC$ circuit. As incident photons break Cooper pairs, the inductance of the superconductor changes, inducing a shift in the resonant frequency that can be read out to determine the amount of power absorbed~\citep{day03}. 

%This focus on a higher frequency range than previous SPT cameras will provide such benefits as improving our understanding of galactic dust and other astrophysical foregrounds as well as extending the redshift range of our SZ galaxy cluster survey to $z > 2$~\citep{anderson22}.

\begin{figure*}
\centering
\includegraphics[width=\columnwidth]{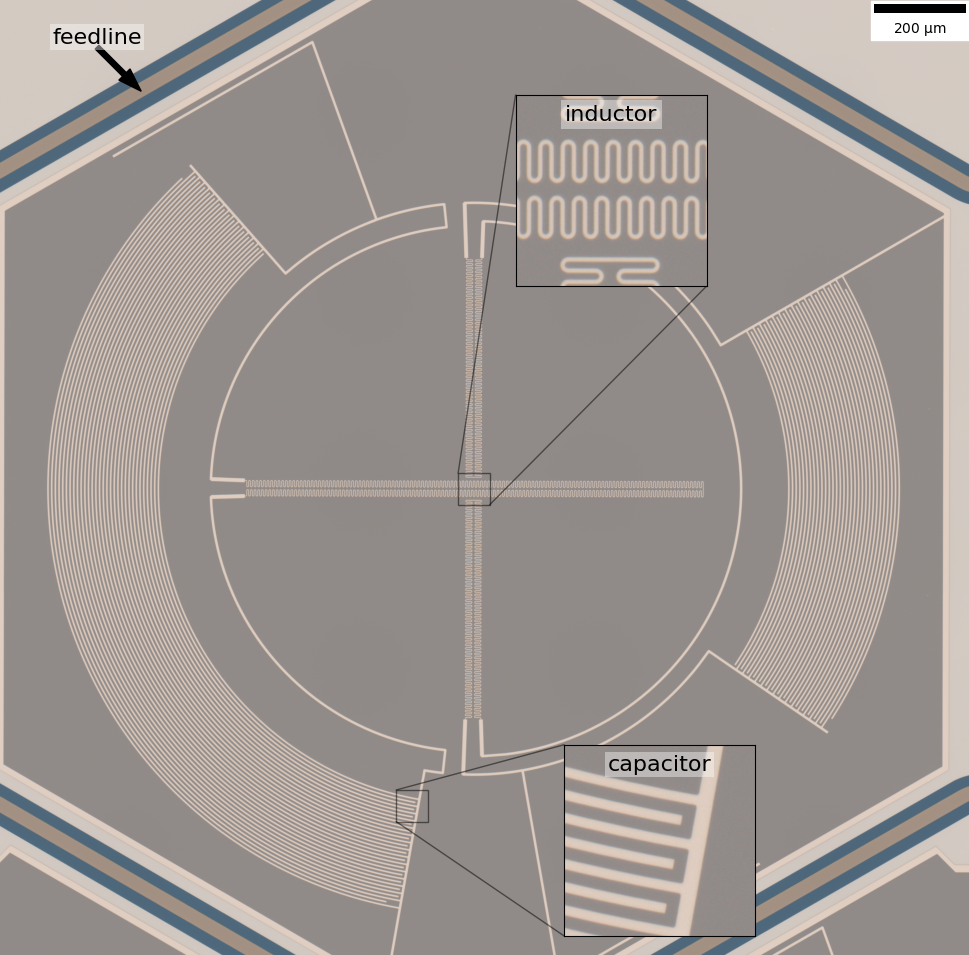}
\caption{Optical image of an SPT-$3$G$+$ $220$ GHz MKID pixel, containing two MKIDs. Insets show detailed views of the inductors (which also act as dual-polarization absorbers), and one of the capacitors (both of which are aluminum). Coupled together, each pair forms an $LC$ resonator. Other materials in this image are niobium for the feedline, silicon dioxide for the feedline surroundings (part of the co-planar waveguide), and silicon for the underlying substrate.}
\label{fig:mkid}
\end{figure*}

A microscopic image of the current iteration of the $220$ GHz pixel design~\citep{dibert23} is shown in Fig.~\ref{fig:mkid}. Each pixel contains two MKIDs, each sensitive to a single orthogonal polarization. The meandered lines forming a cross shape in the center are the inductors and the arcs on the outer edges are the capacitors. While the inductor geometry is identical on every pixel within a given observing band, the angular span of the capacitors and the amount trimmed off their edges are altered to fine tune the resonant frequency of the specific detector. The resonators are capacitively coupled to a co-planar waveguide feedline for microwave readout. Throughout this paper, we will use the term \textit{continuous inductor} to refer to the inductor which is horizontal in Fig.~\ref{fig:mkid}, and \textit{non-continuous inductor} to refer to the one which is vertical and split in the center of the MKID pixel. We will denote the capacitor connected to the continuous inductor (on the left in Fig.~\ref{fig:mkid}) as the \textit{long capacitor}, and the remaining capacitor the \textit{short capacitor}.

\section{Detector imaging}
\label{sec:imaging}
Optical images of the MKIDs are captured on a Zeiss Axio Imager Vario microscope housed at the Center for Nanoscale Materials at Argonne National Laboratory.\footnote{\url{https://cnm.anl.gov/}} The imaging process, including moving the motorized stage to each detector's location, focusing, and capturing the image, is fully automated using the Zeiss ZEN Microscopy Software. Images captured on this instrument are $2752$ pixels wide by $2208$ pixels tall; at our chosen magnification of $20\times$, the scale factor is $0.228$~\textmugreek m per pixel, giving a microscope field of view of about $627$~\textmugreek m wide by $503$~\textmugreek m tall. Because the full detector pixel is approximately $2$~mm across, multiple images (referred to as \textit{tiles}) must be captured and later stitched together before they can be analyzed (see Section \ref{sec:stitching} for stitching details). Images are captured in full color as in Fig.~\ref{fig:mkid}, although only their grayscale information was used for this analysis.

It is vital that the captured images are of high quality, since poor-quality images can compromise the efficacy of the analysis pipeline. For this reason, an exhaustive search of microscope imaging settings was undertaken. Settings that were varied include the magnification, the autofocus strategy, and the overlap between adjacent image tiles. The $20\times$ magnification was chosen in order to balance analysis requirements with computing difficulty; that is, we chose the lowest possible magnification that could meaningfully resolve the smallest features under study in order to minimize the computing resources required to perform the stitching and data analysis. Various microscope autofocus strategies were considered; keeping in mind that the purpose of this pipeline is to save time over traditional cryogenic testing, we found that using Zeiss' ``smart'' autofocus saved significant time over a search of the full focus range, while still yielding in-focus images. We also tested various overlap values between adjacent image tiles (see Section \ref{sec:stitching} for further details). While larger overlaps can provide more information to help the stitching algorithm return high-quality results, they can also lead to more total images captured, increasing the amount of time spent on imaging. We settled on a $30\%$ overlap, yielding seamless stitched images while not increasing the total number of image tiles over a smaller value. Although alternate illumination sources were unable to be installed due to the microscope's multiple users, the exposure time was fixed to a value that provides high contrast in the images\footnote{Since the optimal value can be affected by the level of background light in the room, the exact exposure time varied throughout the data capture period, but was consistently in the $110-130$~ms range.} in order to prevent stitching errors.

\subsection{Stitching details}
\label{sec:stitching}
As previously mentioned, each MKID pixel is larger than the field of view of the imaging microscope at our chosen $20\times$ magnification. In order to fully cover each MKID pixel, $20$ overlapping image tiles are captured in a $5 \times 4$ grid with a $30\%$ overlap between adjacent tiles. A brightness correction is applied to each image tile to account for the slight non-uniform illumination of the microscope's light source. The tiles are then stitched together using a modified version of the Microscopy Image Stitching Tool (MIST) algorithm~\citep{chalfoun17}.

MIST can be summarized in four basic steps (for a thorough walkthrough of the algorithm, see \citet{chalfoun17}):
\begin{itemize}
\item[1)]Estimate translations between adjacent image tiles using the Phase Correlation Method~\citep{kuglin75}, which computes the normalized cross power spectrum of the Fourier-transformed images and equates this to a complex exponential whose phase determines the translation amount.
\item[2)]Refine the estimates using a model of the microscope mechanical stage parameters.
\item[3)]Finalize translations between neighboring tiles down to the single-pixel level by maximizing the normalized cross-correlation $\chi$ of the overlapping region.
\item[4)]Determine the optimal overall set of translations by computing the maximum spanning tree~\citep{erickson19} of a graph where the image tiles are nodes and the $\chi$ values are the edge weights.
\end{itemize}

We find that, on our image data, MIST can occasionally return stitched images with clear stitching errors (likely due to the periodicity in the meandering inductor geometry and the large areas without notable features to help align adjacent tiles). To improve performance, we make the following modifications to the default MIST algorithm, which have the effect of better optimizing the stitching translations at the expense of longer computing time:
\begin{itemize}
\item During the Phase Correlation Method step, default MIST selects translations corresponding to the two maximum values of the Phase Correlation Matrix $P$. These two chosen candidates are often adjacent pixels on the same peak. We instead look for the two highest \textit{local} peaks, requiring that they be at least $10$ pixels apart from one another, in order to better sample the space of possible translations.
\item When finalizing translations, default MIST performs a hill-climbing algorithm, searching for single-pixel-shifted translations between adjacent image tiles that yield larger values of $\chi$ than the original translation. We instead compute $\chi$ over a $44 \times 44$ pixel region; while this leads to the computation of $\chi$ for unnecessary translation values, it ensures that we are finding the true optimal translation rather than a local maximum.
\item During the computation of the maximum spanning tree, default MIST ignores the fact that new translations with un-computed $\chi$ values are introduced for a given spanning tree. This can lead to final images with clear stitching errors, which the image analysis pipeline may falsely flag as detector defects. We use an alternative maximum spanning tree, where the weight for each tree is not the sum of the edge weights (i.e., $\chi$ values) for only the chosen set of translations; instead, the full set of $\chi$ values for every tile overlap is computed and summed together to weight that particular tree. Ideally, this new weight would be computed for every possible spanning tree; in practice, it is restrictively slow to do so for each of the $\mathcal{O}\left( 4{,}000{,}000 \right)$ possibilities for each grid of $5 \times 4$ image tiles. Instead, we examine the $100{,}000$ trees with the highest weights in the default MIST method and choose to use the set of translations corresponding to the highest new weight.
\end{itemize}
Together, these changes drastically increase the quality of the image stitching, yielding a clean output image on which to run the analysis pipeline. For access to a Python implementation of this modified version of MIST (based on the existing \texttt{m2stitch} Python implementation of MIST by Y.~T.~Fukai), see Ref.~\citenum{ferguson24}.

\section{Image analysis pipeline}
\label{sec:pipeline}
The image analysis pipeline has two goals. First it attempts to predict the yield of a chip or wafer by searching for defects that would cause detectors to be inoperable (Section \ref{sec:defect-search}). Then it measures the width of the conducting lines along the inductors and capacitors in order to search for detectors whose resonant frequencies are different than designed (Section \ref{sec:width-measurement}). In order to perform both of these, the pipeline must first determine the location and orientation of the MKIDs in the image.

\subsection{Location and orientation determination}
\begin{figure*}
\centering
\includegraphics[width=0.82\columnwidth]{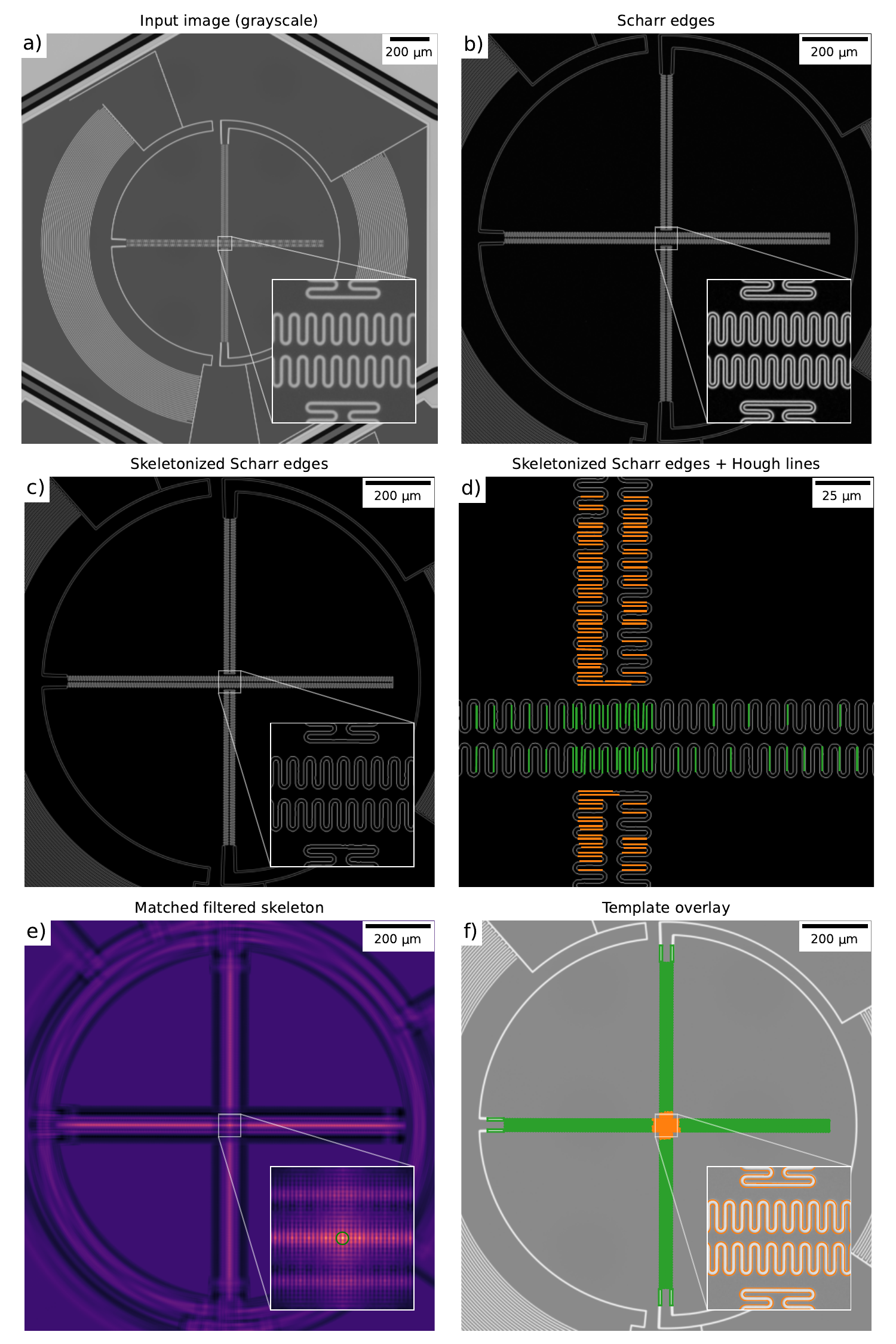}
\caption{Location/orientation determination for MKID pixels in an image. a) The original (stitched) input image is converted to grayscale. b) A Scharr edge detection filter is applied to the image. c) The Scharr edges are skeletonized to become a single pixel wide (the lines have been artificially widened in the zoomed-out view for visual purposes, but are left a single pixel wide in the inset). d) In a small region near the center of the MKID pixel, probabilistic Hough lines are calculated from the skeleton. The lines are split into two groups (represented here by the two line colors) using a $k$-means clustering algorithm in order to determine the orientation of the MKID pixel up to a $90$ degree modulus. e) For each remaining orientation, a template of the inductor skeleton is created from the detector design file, and a small cutout of the center of said template is matched filtered with the skeletonized image to determine the precise location and orientation of the MKID pixel. f) The best-fit template (green) and the small region used in the matched filter (orange) overlaid on the original image.}
\label{fig:MKID-analysis}
\end{figure*}

The pipeline focuses first on the inductors (the two orthogonal meandering lines shown in Fig.~\ref{fig:mkid}), using foreknowledge about their geometry to find the general location and orientation of the MKIDs in the image. The image is converted to grayscale and smoothed using a Gaussian filter with a standard deviation of $1$ pixel (Fig.~\ref{fig:MKID-analysis}a). A Scharr edge detector~\citep{scharr00} is applied (Fig.~\ref{fig:MKID-analysis}b), and the edges are binarized; pixels with values above $25\%$ of the maximum are set to one and all others are set to zero. The binarized array is skeletonized so that all edges are a single pixel wide, and isolated objects containing less than $10$ total pixels are removed in an attempt to reduce the effect of any dust or other small objects that may have been on the detector during imaging (Fig.~\ref{fig:MKID-analysis}c).

Next we perform a probabilistic Hough transform on a small region cut out of the center of the image (Fig.~\ref{fig:MKID-analysis}d). This process is used for line detection and returns a set of line segments corresponding to straight edges in the input skeletonized image. By examining the angles of the returned line segments, we can determine the orientation angle of the MKID pixel modulo $90$ degrees. For each of the four remaining orientation options, we construct a template of the inductor edges from the fabrication design file and match a cutout of the center of this template to the image skeleton. The correct orientation is the one whose matched-filtered output array contains the highest-valued pixel; the location of this pixel also allows us to determine the location of the center of the MKID pixel\footnote{Note the degenerate terminology here. \textit{Pixel} by itself refers to the individual row and column of the image array, while \textit{MKID pixel} refers to the hexagonal area on the detector chip that contains two orthogonal MKIDs.} (Fig.~\ref{fig:MKID-analysis}e, shown for the orientation with the highest maximum value).

Because the area under study is large, a small inaccuracy in angle will lead to prominent issues in the performance of the pipeline far from the center of the MKID pixel. Thus, we refine the orientation angle by constructing templates for $100$ angles uniformly spanning the range $\left[ \theta_{i} - 1^{\circ}, \theta_{i} + 1^{\circ} \right]$ for the initial orientation angle $\theta_{i}$. The angle of the template that has the highest normalized cross-correlation with the image is taken as the final orientation angle moving forward. Note that in this step, we do not compare a template of the inductor edges with the skeleton as before, but rather compare a filled-in template with the Otsu-binarized input image~\citep{otsu79} (shown later in Fig.~\ref{fig:width-boxes}). This gives more accurate angles by penalizing misaligned angles more. The result is shown in Fig.~\ref{fig:MKID-analysis}f with the template lines overlaid on the original image (the orange region in the center of the template is the cutout used for the matched-filter in step e).

%\addtocounter{figure}{-1}
%\begin{figure}
%  \caption{(Caption continued from previous page.) Location/orientation determination for MKID pixels in an image. a) The original (stitched) input image is converted to grayscale. b) A Scharr edge detection filter is applied to the image. c) The Scharr edges are skeletonized to become a single pixel wide (note that the lines have been artificially widened in this figure for visual purposes). d) In a small region near the center of the MKID pixel, probabilistic Hough lines are calculated from the skeleton. The lines are divided into two groups to determine the orientation of the MKID pixel up to a $90$ degree modulus. e) For each remaining orientation, a template of the inductor skeleton is created from the detector design file, and a small cutout of the center of said template is matched filtered with the skeletonized image to determine the precise location and orientation of the MKID pixel. f) The best-fit template (green) and the small region used in the matched filter (orange) overlaid on the original image.}
%\end{figure}

\subsection{Defect search}
\label{sec:defect-search}
\begin{figure*}
\centering
\includegraphics[width=0.8\columnwidth]{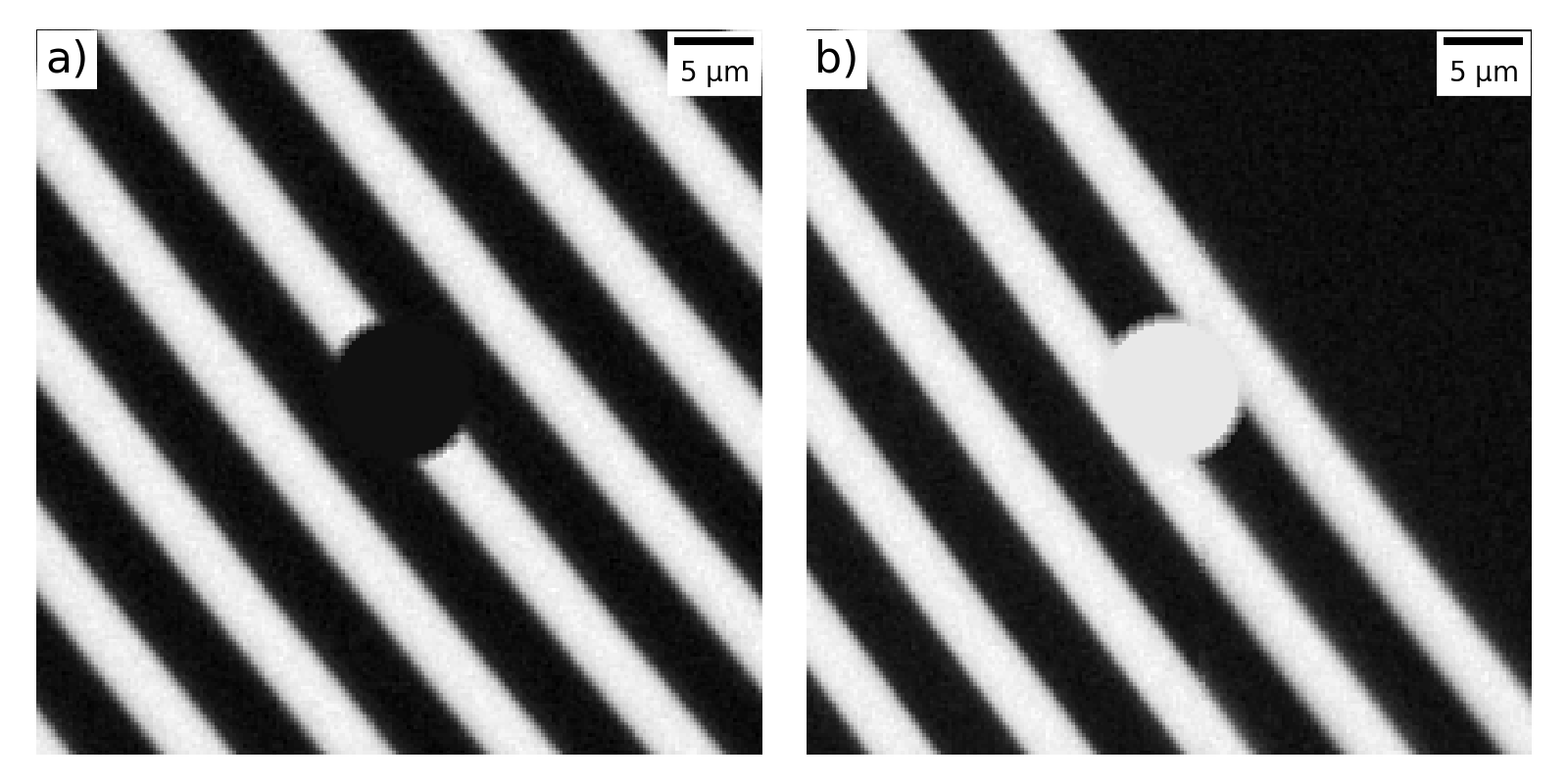}
\caption{Simulated a) subtraction-type and b) addition-type defects along the capacitor conducting lines. For details on the simulation pipeline, see Section \ref{sec:simulations}.}
\label{fig:sim-defects}
\end{figure*}

We search for defects along the inductor and capacitor to estimate detector performance. Defects can take two forms: they can either be a \textit{subtraction}-type defect (i.e., an extra removal of conducting material during fabrication) or an \textit{addition}-type defect (i.e., an extra deposition of conducting material during fabrication). See Fig.~\ref{fig:sim-defects} for a simulated example of each. In other words, we are looking for areas where the fabricated detector differs from its design. Because we are searching for differences, one might think we can simply subtract the binarized image from a template image created in the same location and orientation and look for regions of clustered non-zero pixels. However, the fact that fabricated detectors often have line widths that differ from their design-specified values means there would be a large number of nonzero pixels in a difference array even for an otherwise defect-free detector. In addition, sub-pixel offsets in the location of the detector in the image as well as in the stitching tile alignment can lead the template image to be non-identical to the binarized input image even for a perfectly fabricated detector.

Instead, we use a \textit{flood-fill} algorithm to search for defects. This is equivalent to the paint-bucket tool that comes standard in most image-editing software tools. The flood fill begins at a specified seed location and sets every adjacent pixel of the same value to a specified new value until it runs into a boundary with a different value than that at the seed location. We set one such seed location for each inductor on an MKID pixel. By counting how much of the inductor or capacitor is filled in by this process, we can learn about the presence of defects. While this does not allow us to detect minor defects such as small etches on the side of a conducting line, the method excels at detecting the broken lines that would create an open circuit and cause a detector to be inoperable.

\begin{figure*}
\centering
\includegraphics[width=\columnwidth]{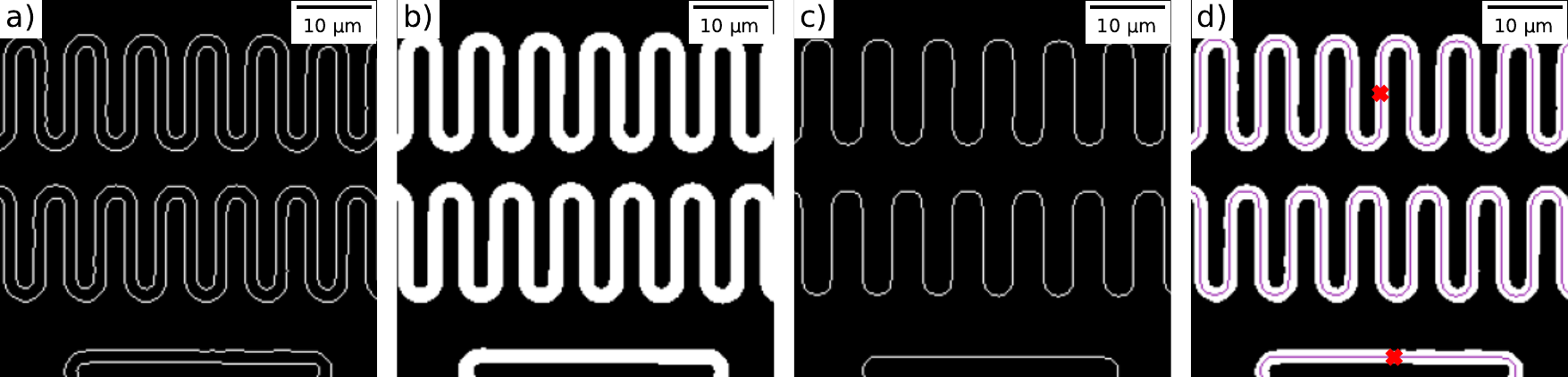}
\caption{a) The skeletonized edges of the input image. b) The inductor template constructed from the design file. c) The skeletonized template. d) The flood-filled image. The skeletonized template, shown in magenta, gives the pixels along which to count the fill fraction. The red exes give the flood fill seed locations.}
\label{fig:example-flood-fill}
\end{figure*}

To perform this accounting, we first skeletonize the inductor template array that was created from the design file during the previous step. This gives a path along the center of the meander lines that we can use to measure the fill fraction. This process is demonstrated visually in Fig.~\ref{fig:example-flood-fill}. A detector is flagged as having a broken line (and therefore being inoperable) if less than $99\%$ of the pixels along this path are filled in. Note that this part of the pipeline is only sensitive to subtraction-type defects on the inductors; addition-type defects, which would change the inductance (and therefore the resonant frequency of the MKID) but not affect operability, will not cause an effect on which the pipeline can flag.

\begin{figure*}
\centering
\includegraphics[width=\columnwidth]{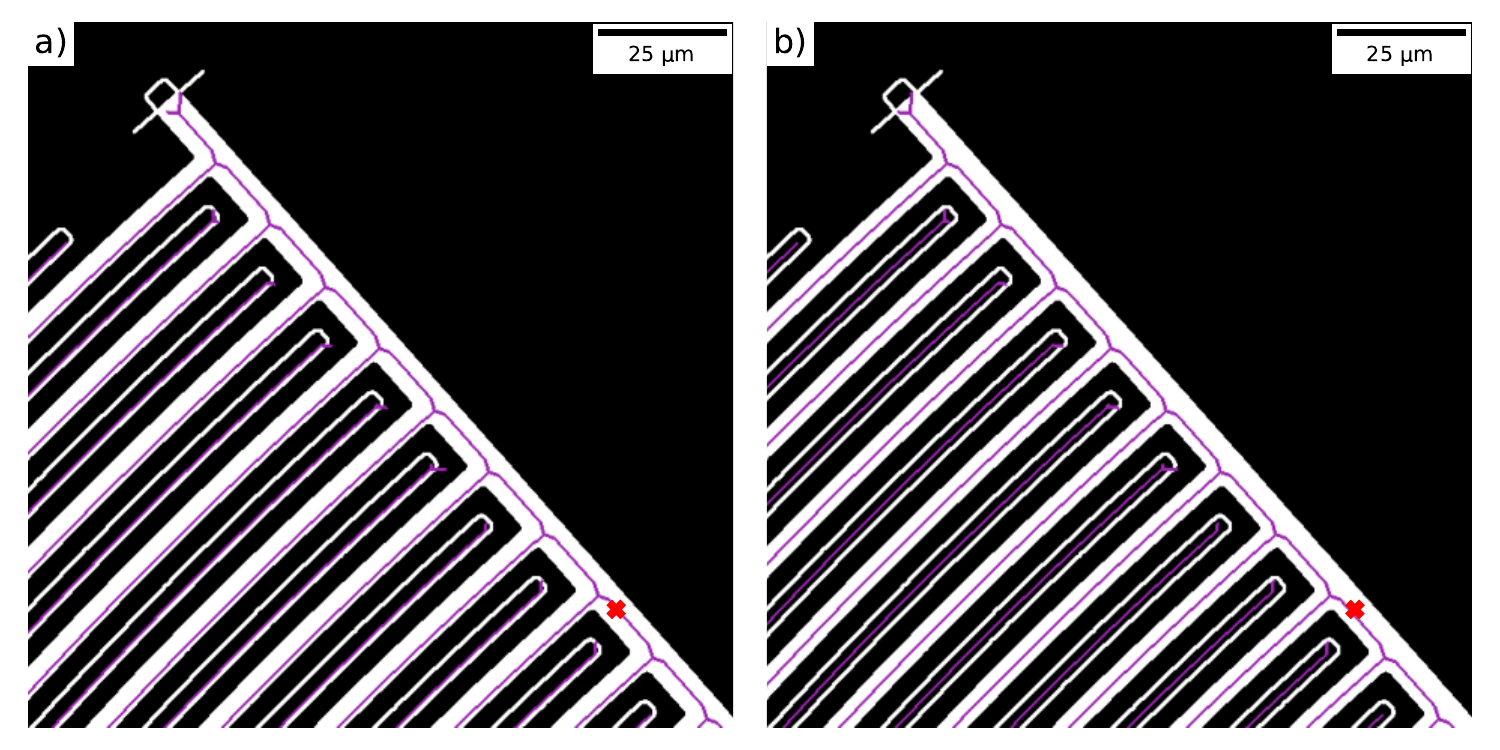}
\caption{The flood-filled capacitor with the a) originally located skeletonized template and b) location-adjusted skeletonized template, both in magenta. The red exes mark the seed location for the flood fill. After adjustment, the skeletonized template path along which flood fill counting is performed is more centrally located. In this figure, all single-pixel-thick paths have been artificially thickened for visibility.}
\label{fig:cap-realignment}
\end{figure*}

The capacitors must be treated differently due to their two electrically isolated, interdigitated legs. Each leg is seeded for the flood fill separately, and the percentage of filled-in pixels is counted for both the seeded and the unseeded leg. Once again, we count along the path defined by the skeletonized template image. The templates tend to be misaligned by ${\sim}1$~\textmugreek m (${\sim}5$ pixels) when using the MKID pixel center calculated earlier in the pipeline. To correct for this, we compute the normalized cross-correlation between the binarized image and the template image in a $20 \times 20$ pixel square centered on the original location and use the location at which $\chi$ assumes its maximum value. The effect of this realignment is shown in Fig.~\ref{fig:cap-realignment}. We also find that increasing the radius of the capacitor arcs by $1.48$ ($1.04$)~\textmugreek m for the long (short) capacitor improves the alignment, yielding better performance during the flood-fill accounting.

\begin{figure*}
\centering
\includegraphics[width=\columnwidth]{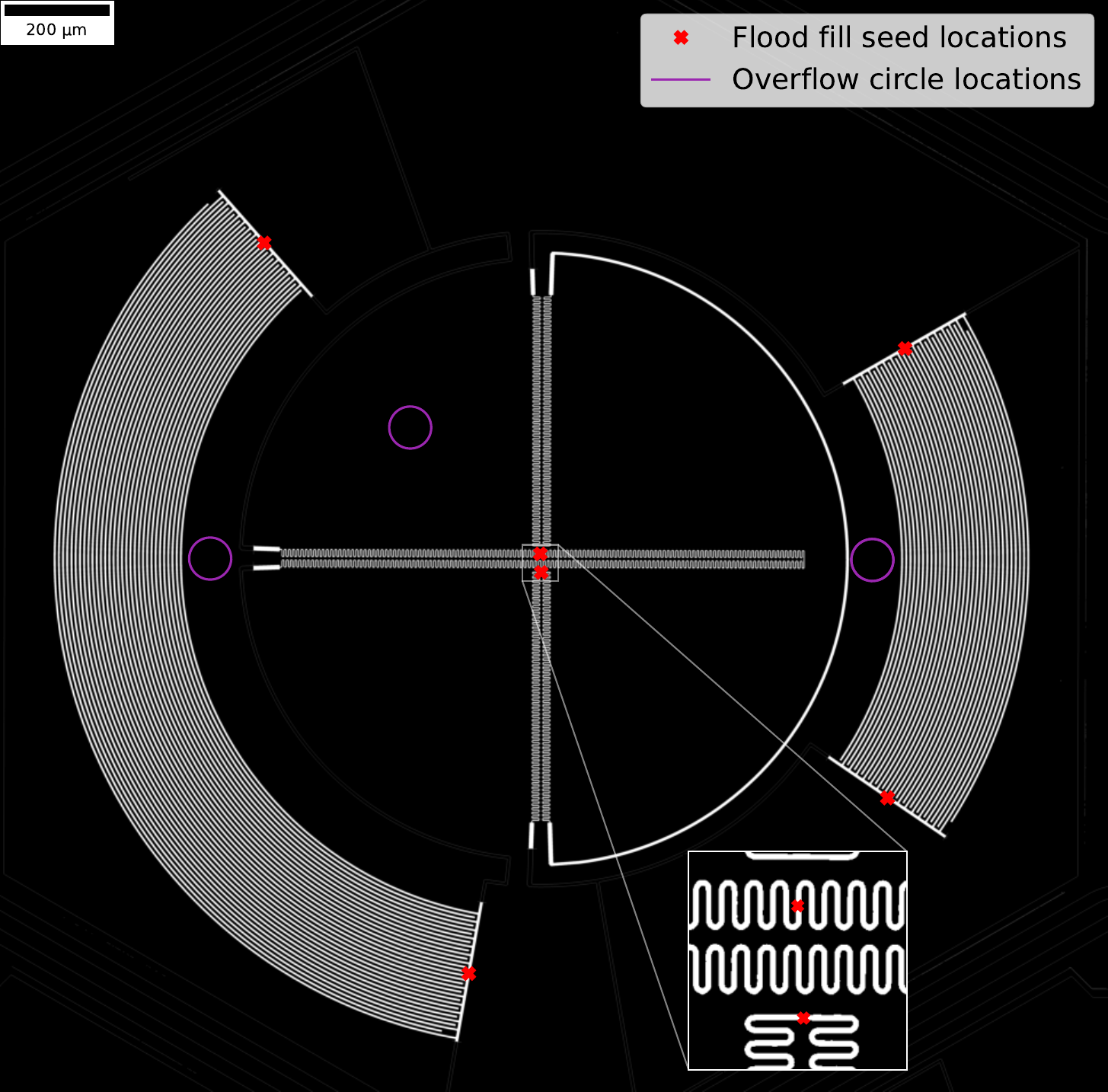}
\caption{An MIKD pixel as the pipeline sees it after the flood-filling algorithm. The red exes are the seed locations for the flood fill (see inset), and the magenta circles denote the areas that are measured for flood fill overflow (described in Section \ref{sec:success-checks}).}
\label{fig:flood-fill}
\end{figure*}

In the capacitors, a broken line does not imply an inoperable detector; rather, it will simply change the capacitance of the capacitor (and therefore the resonant frequency of the detector). However, an addition-type defect that creates a bridge between the two legs will electrically short the capacitor and create an inoperable detector. It is for this reason that we count the percentage of filled-in pixels for both legs while seeding only one at a time. A capacitor with a seeded leg that is less than $99.9\%$ filled will be flagged as having a broken line, while an unseeded leg that is greater than $5\%$ filled will cause the capacitor to be flagged as inoperable. The threshold for a break is higher than in the inductor case because the capacitor lines take up more area, allowing a break the same distance from the end to cause fractionally less of the leg to be unfilled. While the $5\%$ threshold to flag a break may seem unreasonably high, cases with actual bridges tend to cause this quantity to be near $100\%$, so the algorithm is insensitive to the exact threshold. A flood-filled image for an MKID pixel with no breaks, highlighting the seed locations, is shown in Fig.~\ref{fig:flood-fill}.

\subsection{Measuring line width}
\label{sec:width-measurement}
\begin{figure*}
\centering
\includegraphics[width=\columnwidth]{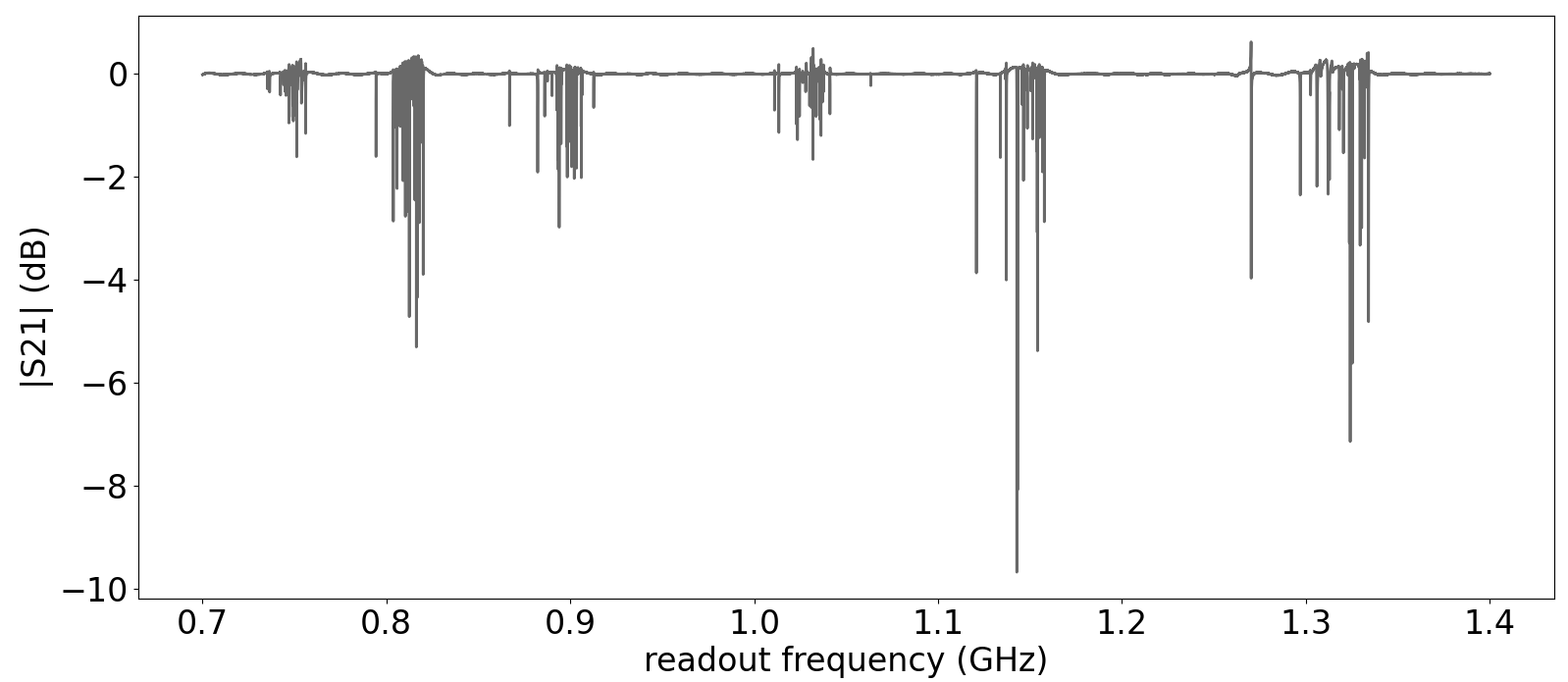}
\caption{The measured frequency response of the first of the two chips under study. $|S_{21}|$ is the power measured out relative to the power put in. On an MKID's resonant frequency, its impedance increases, causing the dips in power. A high-pass filter has been applied to these data to reduce large-scale variation in baseline frequency response across the readout bandwidth. This figure was produced from work described in \citet{dibert23}}
\label{fig:freq-sweep}
\end{figure*}

The other major output of the pipeline is a measurement of the width of the inductor and capacitor conducting lines. As this quantity varies from its specified value, the inductance and capacitance change, inducing a shift in the resonant frequency of the detector. The SPT-$3$G$+$ detectors are designed with resonant frequencies spanning from about $0.73$ to $1.33$ GHz. For the specific chips under study here, there are six distinct frequency banks, each containing $27$ MKIDs spaced $500$ kHz apart. The measured frequency response for one chip is shown in Fig.~\ref{fig:freq-sweep}. Preliminary models indicate that a $10\%$ shift in line width will induce a ${\sim}4\%$ shift in inductance, in turn inducing a ${\sim}2\%$ shift in resonant frequency; thus, even small relative differences in line width have the potential to meaningfully increase the resonance scatter \citep{li22}. Increasing the line width is expected to decrease inductance (increase capacitance), in turn increasing (decreasing) the resonant frequency. However, the exact relationship is complex and a full parameterization of it is the subject of future work.

\begin{figure*}
\centering
\includegraphics[width=\columnwidth]{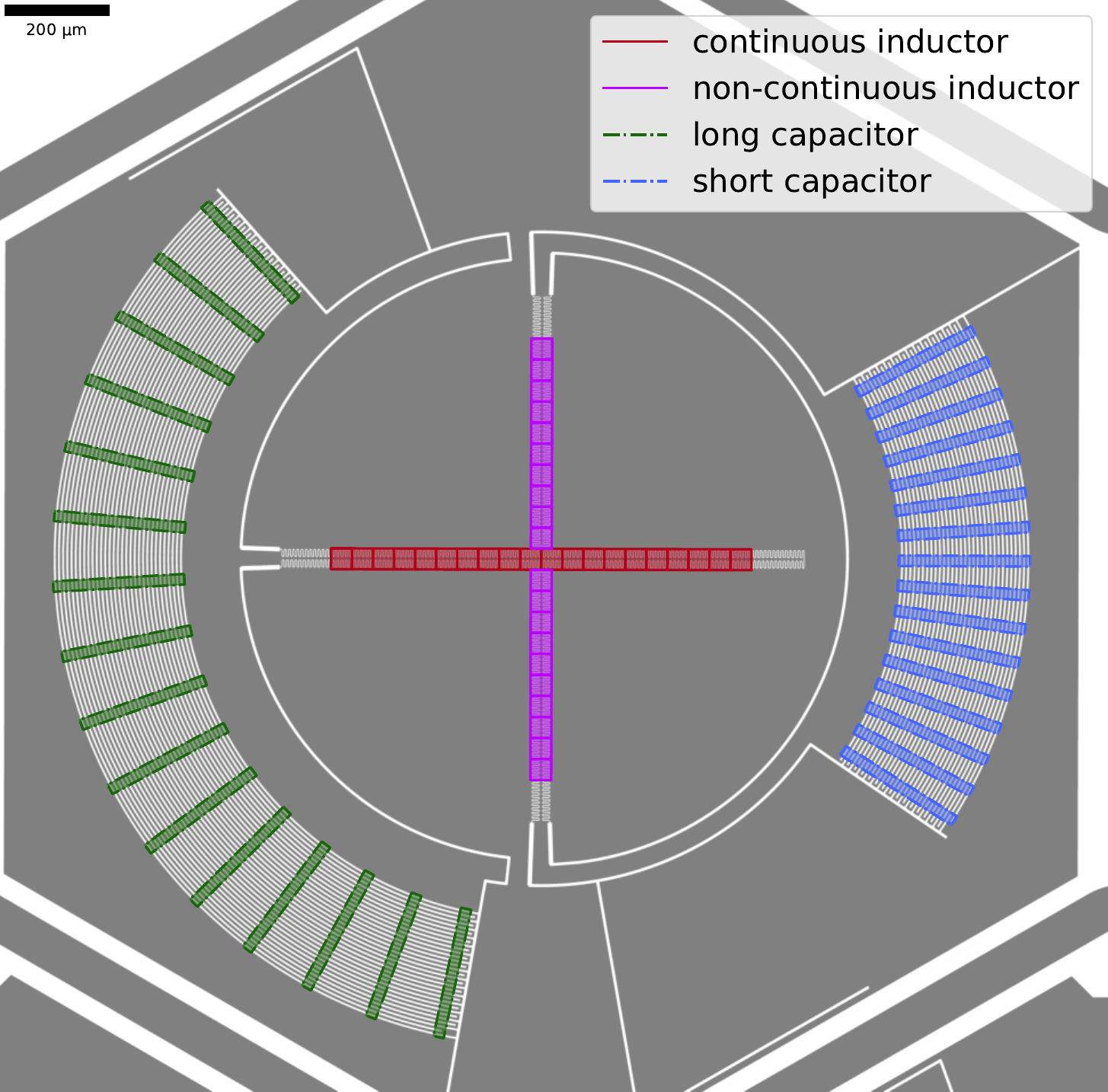}
\caption{The locations of the regions used to measure the width of the conducting inductor and capacitor lines, overlaid on the Otsu-binarized input image. The \textit{continuous inductor} is the horizontal one covered by the red regions, and forms a circuit with the \textit{long capacitor}, on the left and covered by the green regions. The \textit{non-continuous inductor} is the vertical one covered by the magenta regions, and forms a circuit with the \textit{short capacitor}, on the right and covered by the blue regions. Note that the regions always cover the full angular span of the capacitors, even as said span changes between different MKID pixels.}
\label{fig:width-boxes}
\end{figure*}

The line width is measured separately for each inductor and capacitor on an MKID pixel. We specify rectangular regions that span the object under study in the Otsu-binarized images. The number of pixels with value $1$ in each region in the binarized image (i.e., the area of the inductor/capacitor in the box after thresholding) is divided by the number of pixels with value $1$ in the same region in the template image. We define the quantity $\delta w \equiv w_{\text{measured}} / w_{\text{original}} - 1$ to represent the fractional shift in line width (note that although we are dividing areas and not distances, the path length is identical in both the original and measured cases and cancels out, leaving only the ratio of widths). With this definition, a negative value of $\delta w$ corresponds to a line that is less wide than it was designed to be, while a positive value of $\delta w$ corresponds to a line that is wider than designed. This measurement is done for a number of small regions over the full area of the inductor/capacitor, and the final measurement is taken as the mean and standard error of the individual measurements. The location of the regions used is shown over the Otsu-binarized image in Fig.~\ref{fig:width-boxes}.

\subsection{Pipeline success checks}
\label{sec:success-checks}
Occasionally, something will go wrong during the image analysis. The most common failure mode is a ``weak edge,'' an edge where the Scharr filter returns abnormally low values that fall below the $25\%$ binarization threshold mentioned above. This causes a gap in the skeleton, which leads to an overflow during the flood fill algorithm.

To catch the occasional case where this happens, we sample pixels in circular regions located in pre-chosen areas where we expect there to be no filled-in pixels of the flood-filled images. The locations of these regions are shown by the circles in Fig.~\ref{fig:flood-fill}. We count the number of filled-in pixels inside of these regions and flag the pipeline's results as \textit{questionable} if a region has greater than $5\%$ of its pixels filled in. This threshold is high enough to not trigger on the occasional piece of dust or other detritus that causes some filled-in pixels inside of the region, but low enough to trigger on something that has actually overflowed, as the value in these cases is at or near $100\%$.

\section{Simulation framework}
\label{sec:simulations}
We simulate defects in images to test the purity and completeness of this pipeline. By analyzing images where the ``truth'' of whether there is a defect that the pipeline should detect is known in advance, we can compare said truth to the results to better understand pipeline performance.

\subsection{Defect simulation}
\label{sec:defect-sims}
To create simulated images, we start with a clean input image that has been verified to be free of defects. This ensures that defects found in simulated images are interpreted correctly. Cleanliness of the input image is verified both through visual inspection and application of the pipeline.

We specify five quantities when creating a simulated image: the input image to use, the number of defects to introduce $N_{\text{def}}$, the size of the defects $r_{\text{def}}$, the location of the defects (on the inductors or on the capacitors), and the type of defect (subtraction-type or addition-type). For simplicity, we only introduce circular defects. While real fabrication defects are generally expected to be more geometrically complex, circular defects nonetheless allow us to test the desired aspects of the pipeline; namely, determining the success rate of defect finding and probing any unexpected behavior. Because the input image has already been run through the pipeline, we also know the location and orientation of the inductors/capacitors, allowing us to restrict the defect placement to these regions. Note that defects are not placed only on the conducting lines; they are allowed to be seeded anywhere in the general areas spanned by these objects. In other words, we simply exclude the large open areas where a defect would not affect performance or even be visible to the pipeline. Within the specified regions, a random pixel is chosen as the center of the defect. Subtraction-type (addition-type) defects are given a value equal to the median value of all pixels belonging to the background substrate (the conducting lines) in the image.

Once a defect is placed, we need to determine the ``truth'' of that defect. We define a \textit{true defect} as one that is serious enough that we expect the pipeline to flag it (i.e., it is either a complete line break anywhere or a complete bridge between opposite legs of the capacitor), and a \textit{false defect} as one that does not meet these criteria. To determine a defect's truth value, we find contours on the original images along the grayscale value halfway between the values assigned to the two defect types. This effectively gives one contour line per edge. By counting how many of these contours are broken by the contour of the defect, we are able to determine the truth value of that defect. On the inductors, a subtraction-type defect that breaks at least two contours is considered true; although addition-type defects on the inductors will affect the detector resonance, in simulations we always label these as false defects since the pipeline is not sensitive to them. On the capacitors, a defect that is of the opposite type as its seed location (that is, a subtraction-type defect seeded on the background substrate or an addition-type defect seeded on the conducting lines) that breaks at least two contours is considered true, while a defect that is of the same type as its seed location needs to break at least three contours to be considered true.

The following inputs are used to generate our suite of simulated images:
\begin{itemize}
\item $11$ clean input images
\item $N_{\text{def}} \in \left[ 1,2,3,4 \right]$
\item $r_{\text{def}} \in \left[ 10,15,20,25 \right]$ (note: this is the radius of the circular defect in pixels)
\item Defect location $\in \left[ \text{inductor, capacitor} \right]$
\item Defect type $\in \left[ \text{subtraction, addition} \right]$
\end{itemize}
Altogether, this gives $704$ simulated images to run through the pipeline.

\subsection{Simulation results}
The $704$ simulated images are run through the analysis pipeline as though they are real data images (with the one exception that the simulated images are saved to disk in grayscale, so they do not need to be converted to grayscale at the start of the pipeline). Although we flag each detector on an MKID pixel separately when running on real data, on the simulated images we simply flag the whole image (both in the truth determination and when running the image through the pipeline).

\begin{figure*}
\centering
\includegraphics[width=0.9\columnwidth]{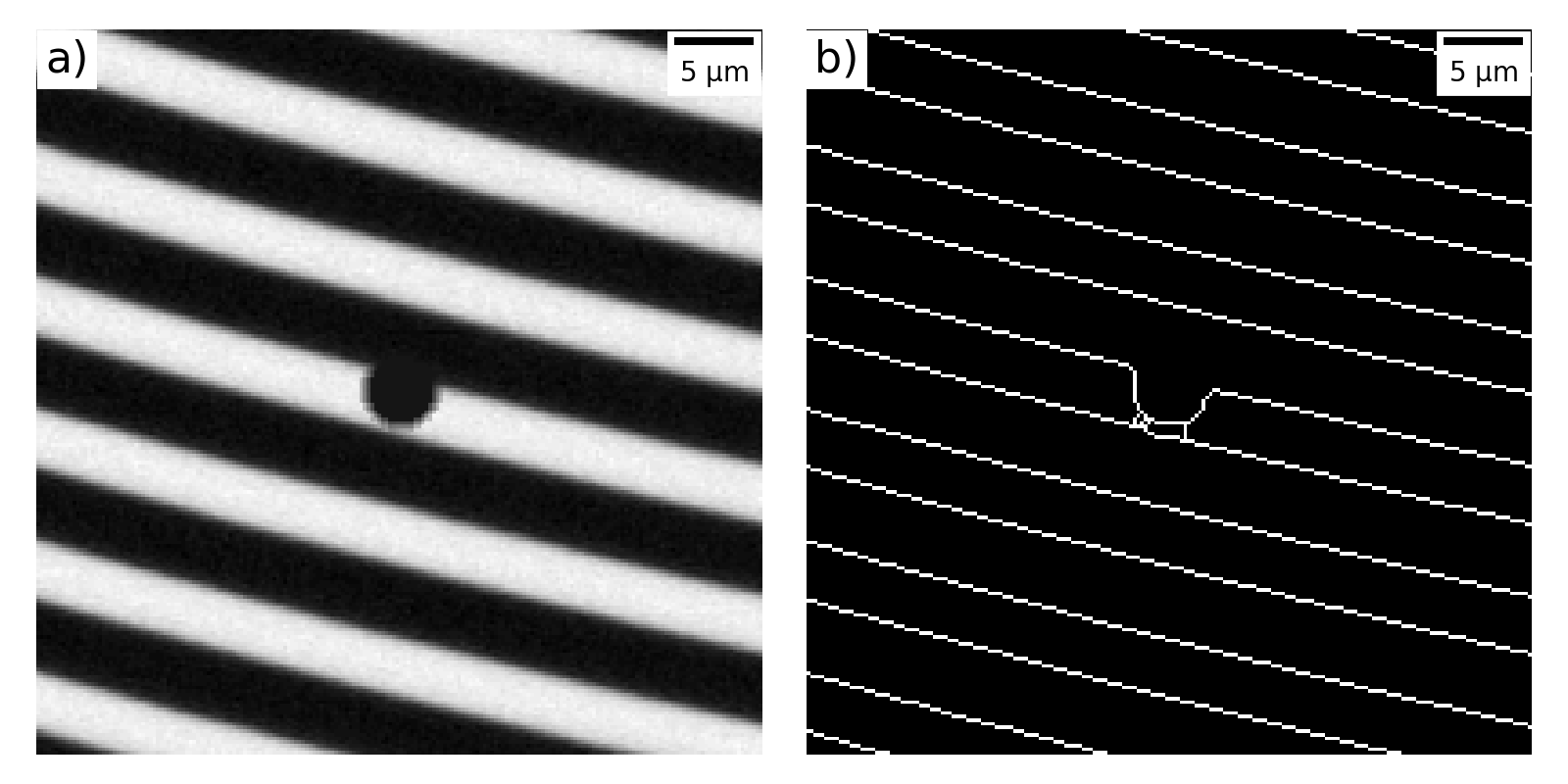}
\caption{a) An example of a simulated subtraction-type defect that does not actually break a line, but eats enough of the line away that the analysis pipeline flags it as a complete break. b) The skeletonized array around this simulated defect, showing the gap-bridging line that initially caused the pipeline to flag cases such as these as false positive defect detections.}
\label{fig:almost-defect}
\end{figure*}

We characterize the results of the simulations by the number of false negatives (images that were found to have a true defect but the analysis pipeline did not flag) and false positives (images that were flagged by the pipeline but did not contain any true defects). In total, we find $5$ false negatives and $51$ false positives (along with $193$ true negatives and $455$ true positives). Upon visual inspection of the defects introduced in these images, there are some clear trends. Of the false negatives, three are located on the inductors. In each case, although there is a true break, there is also an addition-type defect bridging across the break somewhere else on the same inductor. With the flood fill circumnavigating the break, enough of the inductor is filled in that the image is not flagged by the pipeline. Among the false positives, the vast majority include subtraction-type defects that come close to fully breaking a line but do not do so completely (see Fig.~\ref{fig:almost-defect} for an example). The remaining line is thin enough that the skeletonization process connects the two remaining edges, creating a false break that cuts off the flood fill (note that, due to the specifics of the edge detection transform, addition-type defects that nearly-but-not-fully bridge the capacitor gap appear to not create false positives in the same way). Though current may still be able to run through these lines, defects such as this will likely still affect the performance of the detector and so it is desirable that the pipeline is flagging such geometries. Nevertheless, this is an unanticipated behavior and so to fully characterize the pipeline, we must determine how thin a line can be before it is flagged by the pipeline.

In order to do so, we create a new set of simulated images, each with only a single defect but keeping all of the other input parameters the same. We generate $5$ simulated images for each combination of input parameters, yielding a total of $880$ new, single-defect simulations. In these simulations, since we are looking specifically for false positives, we do not create any images with true defects. For each defect, we measure the distance between the defect edge and the closest unbroken contour. These images are then run through the analysis pipeline, after which $97$ images are flagged as containing defects. However, upon visual inspection of these images, $22$ are found to actually contain true defects. This is possible because in certain edge cases, the defect truth determination criteria described in Section \ref{sec:defect-sims} can return an incorrect truth value. Specifically, this happens when the defect is at the edge of a capacitor (because the logic expects the presence of a contour that is not there in reality), or when the defect is a subtraction-type defect near an existing edge (because this will weaken the edge enough for the skeletonization to fail and the flood fill to spill out). $22$ of the $97$ ``false'' defects fell into one of these two categories; that is, they were actually defects that we expect (and want) the pipeline to flag. Because of this, we do not include them in the calculation of minimum line thinness.

\begin{figure*}
\centering
\includegraphics[width=0.75\columnwidth]{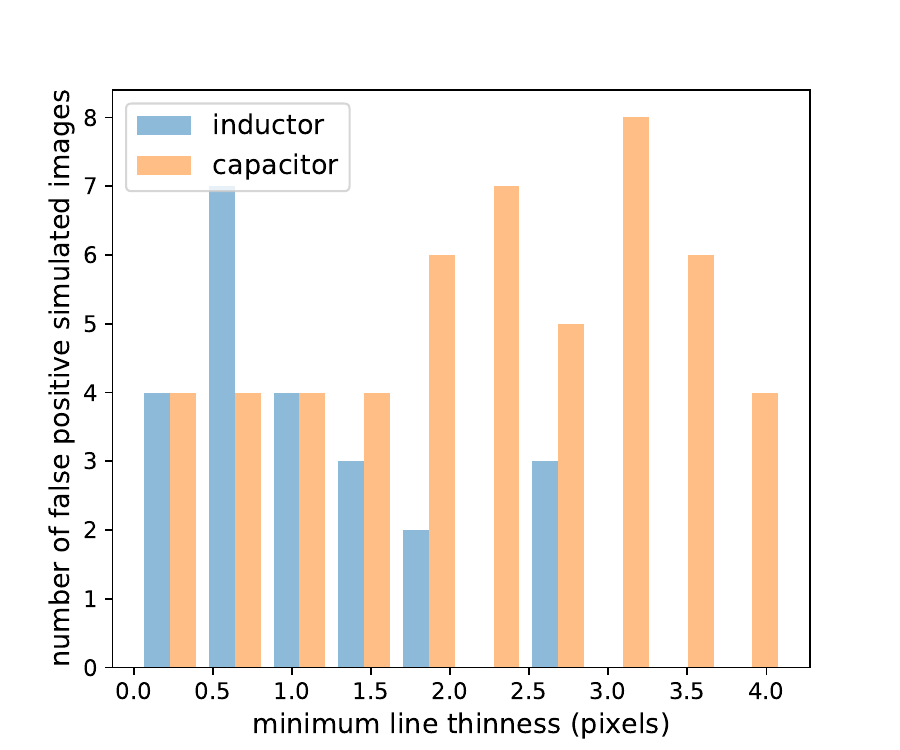}
\caption{Distribution of minimum line thinness for simulated false positive images, split by whether the simulated defect was on the inductors or the capacitors.}
\label{fig:thinness-dist}
\end{figure*}

The line thinness distribution for these images is shown in Fig.~\ref{fig:thinness-dist}. It falls off sharply for distances above ${\sim}2.75$ pixels for the inductors, and above ${\sim}4$ pixels for the capacitors. It is unclear why the distribution for inductor-located defects has a different shape than that for capacitor-located defects, but this is assumed to be an effect of the different geometry between the different objects. The largest line thinness for which a false positive was still found is $4.12$ pixels,\footnote{It is possible to calculate these distances with sub-pixel precision because the contour-finding algorithm (part of the \texttt{measure} module in the \texttt{scikit-image} Python package) uses linear interpolation to return real-valued contours, not integer-valued ones.} corresponding to a physical distance of $0.94$~\textmugreek m (for context, the capacitor lines are designed to be $4$~\textmugreek m wide, and the inductor lines $2$~\textmugreek m wide). In other words, the analysis pipeline is unable to distinguish complete line breaks from subtraction-type defects that leave a line intact but with a remaining thinness $\lesssim 0.94$~\textmugreek m.

We then return to our defect truth determination, and reclassify any subtraction-type defect that is one broken contour away from being a true defect as a true defect as long as the unbroken contour is less than $4.12$ pixels away from the defect edge. Once this is done, the number of false positives in our original simulation set changes from $51$ to $5$ (and the number of true positives from $455$ to $501$), giving an overall defect-finding accuracy of $694/704 = 98.6\%$. Along with the $5$ false negatives, this gives a precision equal to the recall of $0.990$. All-in-all, the simulation suite is highly successful, both in stress-testing the analysis pipeline (revealing a quirk in the behavior of which we were previously unaware), as well as in demonstrating that the pipeline has a remarkably high rate of success in correctly classifying whether or not an image contains defects.

\section{Results and discussion}
\label{sec:results}
For the initial study of the pipeline we imaged two prototype chips, each containing $81$ MKID pixels ($162$ MKIDs). Each chip was fabricated as part of a larger wafer and later diced for individual imaging and testing. We will refer the chips using the monikers \textit{chip one} and \textit{chip two}. The MKID pixels on chip one are all oriented in the same direction, while every other MKID pixel on chip two has its inductors rotated $135$ degrees relative to its capacitors. We apply the pipeline described in Section~\ref{sec:pipeline} to search for defects and measure the inductor and capacitor line width for each detector.

\subsection{Defect search}
\begin{figure*}
\centering
\includegraphics[width=0.5\columnwidth]{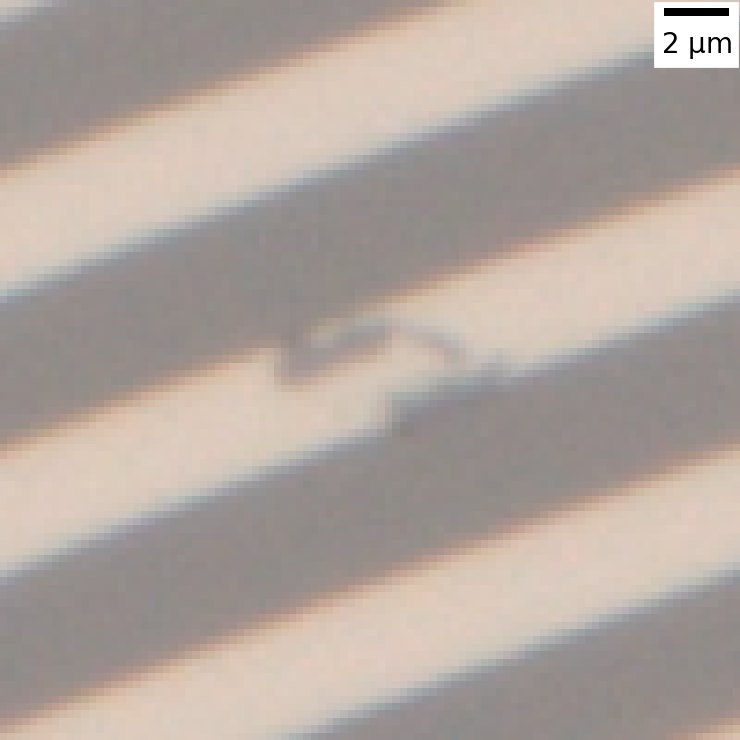}
\caption{The single defect found by the pipeline across the $81$ MKID pixels on chip one, located along one of the capacitor fingers.}
\label{fig:mkid-defect-one}
\end{figure*}

Chip one is remarkably defect-free. The pipeline finds only a single defect over the full set of detectors: an odd-looking line break along one of the capacitor fingers, shown in Fig.~\ref{fig:mkid-defect-one}. This defect is not caught because it causes a part of the capacitor leg to be un-flood-filled, but instead because it weakens the edges of the line enough for the flood fill to overflow and trigger one of the pipeline success checks described in Section~\ref{sec:success-checks}.

\begin{figure*}
\centering
\includegraphics[width=0.5\columnwidth]{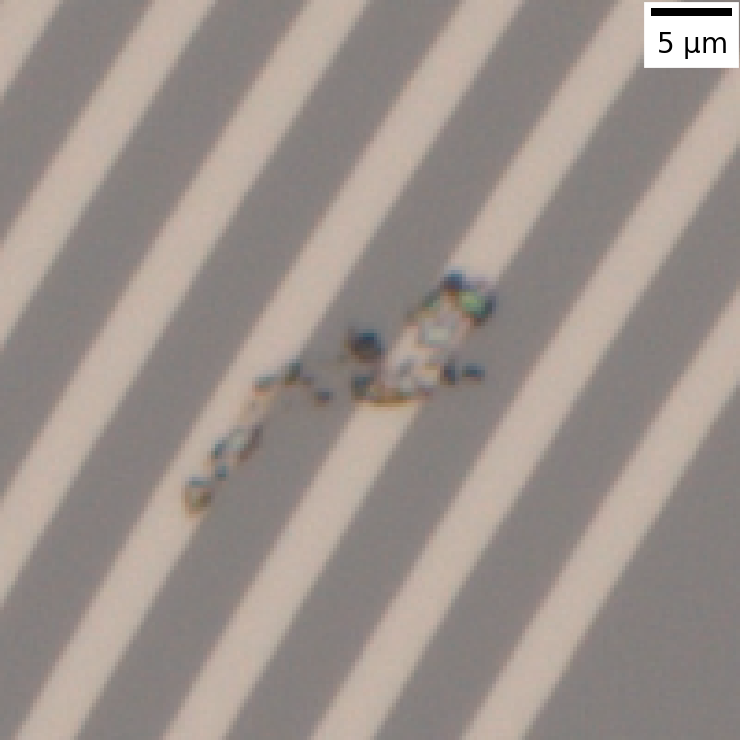}
\caption{An example of the iridescent defects common across chip two, located along one of the capacitor fingers. These appear to be either hardened photoresist or additional dielectric material atop the conducting lines of the detector. Given that neither of these materials are conductive and that they are not actually breaking any conducting lines, we expect detectors containing these types of defects to still function electrically. Other quantities, however, such as their resonant frequency and optical performance, may be impacted.}
\label{fig:mkid-defect-two}
\end{figure*}

On chip two, the pipeline detects a defect of some sort (be it a broken inductor or capacitor line, a bridge between capacitor legs, or a detector where the flood fill overflowed as with the defect on chip one) on $131$ detectors. Upon visual inspection of the defects that were found, most appear as a layer of iridescent material on top of the detector, as shown in Fig.~\ref{fig:mkid-defect-two}. Before the chips are imaged, they are cleaned for two minutes in an ultrasonic acetone bath followed by two minutes in an ultrasonic isopropyl alcohol bath, then blow dried with molecular nitrogen. Therefore, it is likely that these defects are an artifact of the fabrication process for chip two rather than something that made it dirty afterwards. Possible causes for defects with this appearance could be a failure to remove hardened photoresist or a defect in the photoresist causing the deposition of dielectric material where it should not be.

Because the defect on chip one is a line break on the capacitor, we do not expect it to affect the ability of the detector to operate; only the capacitance, and therefore the resonant frequency of the detector, should be affected. Therefore, the pipeline predicts a yield of $100\%$ on this chip. On chip two, the pipeline finds $89$ detectors with explicit defects that are expected to make the detector inoperable (that is, either a broken line in the inductor or a bridged gap in the capacitor). If one assumes that every instance of a flood fill overflowing equates to a line break with a weak edge (as is the case in the capacitor defect found on chip one), then we can add to this set the inductors that had overflowing flood fills, giving $100$ detectors expected to be inoperable.\footnote{The number of detectors on chip two with problems that was reported earlier, $131$, includes broken and overflowing capacitor lines, which are not expected to affect detector operability.} Without additional information, the pipeline predicts a yield for chip two between $38.3\%$ and $45.1\%$, depending on how one chooses to interpret the overflowing detectors. However, neither of the likely defect types mentioned above (hardened photoresist or extra dielectric material) should affect the operability of the detector because they do not break any lines and are not themselves conductive. Nevertheless, it is important to detect defects of this type because it is possible they could affect other aspects of detector performance such as the resonant frequency or the optical efficiency. However, without additional image or cryogenic information, it is difficult to say how many of the found defects are truly caused by broken or bridged lines and how many are caused by additional material on top of the detector.

\subsubsection{Comparison with cryogenic yield measurements}
The yield of each chip is measured by installing the chip in a cryogenic system, sending in a pure tone at a fixed frequency, and measuring the response as this frequency is slowly swept across the full range of expected resonant frequencies. By counting the number of distinct dips in the frequency-dependent transmission (Fig.~\ref{fig:freq-sweep}), one can determine the number of functional detectors. When chip one was tested in a cryogenic system, $143$ detectors were found to be operable, giving a true yield of $88.3\%$. For chip two, $146$ working detectors were found, giving a true yield of $90.1\%$. There are many possible reasons for the discrepancy between the pipeline predictions and these measurements:
\begin{itemize}
\item Detectors that are operable but with small resonances are difficult to distinguish from variability in the baseline.
\item Detectors with overlapping resonant frequencies will only be counted once.
\item Defects can also appear along the lines that connect the inductor and the capacitor. These would impact performance but are not visible to our pipeline. Automated defect searches in this part of the chip geometry are considered for future work.
\item As previously stated, though the pipeline will flag iridescent defects of the kind shown in Fig.~\ref{fig:mkid-defect-two}, these defects are not expected to affect detector operability.
\item Visual information does not capture all the characteristics of the detector, such as the material properties of the superconducting film.
\end{itemize}

%Recall that the purpose of this pipeline is to provide an initial assessment of yield and potential quality to reduce time spent testing poorly performing detectors.

Because the purpose of this analysis technique is the identification of potentially high-yield wafers for further (cryogenic) testing, we neither anticipate nor intend for an exact match between the number of visually identified defects and the number of cryogenically identified defects. Rather, in the expected low-yield regime of detector-wafer production, we intend for the pipeline to separate those wafers that are clearly problematic and can be discarded early, without cryogenic testing, from those that are potentially viable. As such, we intentionally do not worry about exact fault counts, as for example in the second item in the list above.

In this case, the pipeline would recommend that chip one be tested cryogenically; once the $88.3\%$ yield was actually measured, a decision could be made on what to do with the chip. The yield prediction for chip two is low enough that in a true production environment, we would recommend discarding it before cryogenic testing. Though it may seem as if this would cause us to pass over a well-performing chip, a device like chip two would likely be passed over anyways in favor of other, cleaner chips due to the number of other, subtle ways defects of this type could affect detector performance. These kinds of deposits on the wafer would, in particular, be expected to negatively affect optical performance, which was not cryogenically tested; a wafer like this would not be deployed as-is.

\subsection{Line width measurement}
\begin{figure*}
\centering
\includegraphics[width=\columnwidth]{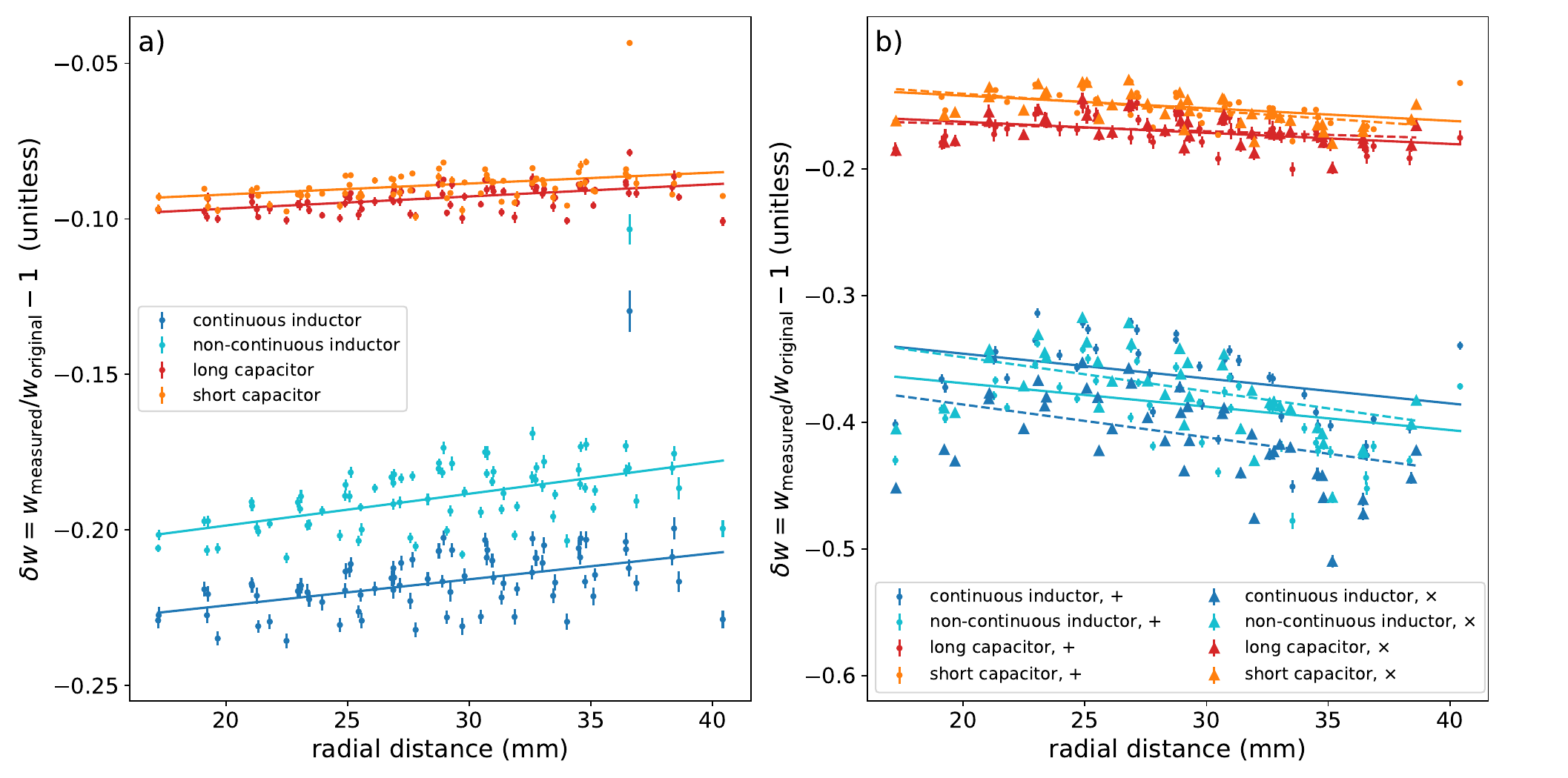}
\caption{The measured width of the conducting lines on a) chip one and b) chip two as a function of the MKID pixel's radial distance from the center of the wafer. A linear trendline has been fit to each population of data points. Note that, because the detectors on chip two are in two different orientations, a separate trendline has been fit to the data points from each orientation as well. The solid trendlines are fit to the detectors with the $+$ orientation (the same orientation as all detectors on chip one, and the circle data points in this figure), while the dashed trendlines are fit to the detectors with the $\times$ orientation (the triangle data points in this figure).}
\label{fig:line-widths}
\end{figure*}

As discussed in Section \ref{sec:width-measurement}, the other main product of the pipeline is a measurement of the capacitor and inductor line width. The measured relative line width of each detector on chip one is plotted as a function of radial distance from the center of the substrate wafer during fabrication in Fig.~\ref{fig:line-widths}a, along with a linear fit to each population. Fig.~\ref{fig:line-widths}b shows the same for chip two, except that separate trendlines have been fit to the data points corresponding to detectors with different inductor orientations ($+$ for inductors of the same orientation as those on chip one and $\times$ for those that have been rotated). Note that the radial distances refer to the \textit{center} of each MKID pixel; sub-MKID-pixel variation is not captured here.

%\begin{figure*}
%\centering
%\includegraphics[width=0.8\columnwidth]{figures/detectors/all_width_vs_radius_chip_two.pdf}
%\caption{The measured width of the conducting lines on chip two as a function of the MKID pixel's radial distance from the center of the wafer. A linear trendline has been fit to each population of data points. Note that, because the detectors on chip two are in two different orientations, a separate trendline has been fit to the data points from each orientation as well. The solid trendlines are fit to the detectors with the $+$ orientation (the same orientation as all detectors on chip one, and the circle data points in this figure), while the dashed trendlines are fit to the detectors with the $\times$ orientation (the triangle data points in this figure).}
%\label{fig:line-widths-two}
%\end{figure*}

There are a few features that warrant discussion.
\begin{itemize}
\item The widths of all lines on both chips are smaller than they were designed to be; that is, $\delta w < 0$.
\begin{itemize}
\item This also varies by chip; the line widths measured on chip two are significantly lower than those measured on chip one.
\end{itemize}
\item While there are nonzero radial trends present in all populations, the direction of these trends is different on the two chips. Several fabrication process steps have a radial symmetry; therefore a radial trend could be introduced during fabrication. It is possible that different steps dominated the introduction of the trends for the two chips, leading to the different trend behavior.
\item While at first glance it may appear that the capacitors lose less width than the inductors do, recall that we measure a ratio of widths. The capacitor lines are designed to be $4$~\textmugreek m wide; the inductor lines meanwhile are only $2$~\textmugreek m wide. Etching away the same amount on each object will cause a larger fractional loss in the inductors than it will in the capacitors.
\item On both chips, the scatter in the inductor data is much greater than in the capacitor data. This is likely also due to the fact that we measure fractional line width and the inductors are designed with smaller line width.
%\item The lines belonging to the broken inductor and short capacitor (which, recall, go together to form a single detector in an MKID pixel) tend to be slightly wider than the lines belonging to the unbroken inductor/long capacitor. This does not appear to be a systematic effect of the measurement process, as it remains true even when the images are artificially rotated before performing the measurement. The exact cause remains unknown.
\item Objects in different orientations have systematically different line widths. This is especially evident on chip two, where each trendline fit to the inductor data exhibits notably unique behavior. Together with resonant frequency measurements of individual detectors on this chip revealing a clear offset between frequencies for detectors on $+$ MKID pixels and $\times$ ones, this provides evidence of a directionality bias in the fabrication process \citep{martsen24}.
\end{itemize}

\begin{figure*}
\centering
\includegraphics[width=\columnwidth]{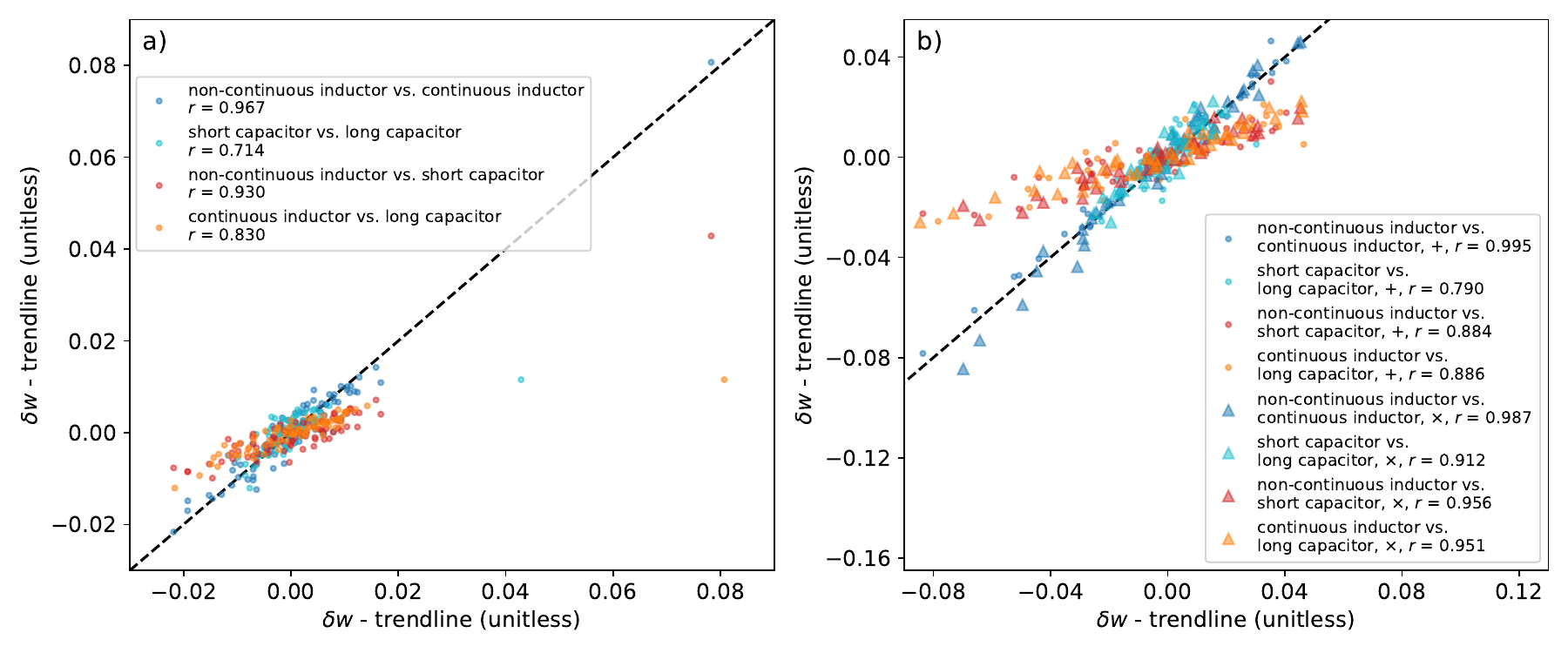}
\caption{A visualization of the line width correlation between the objects within a single MKID pixel on a) chip one and b) chip two, relative to their linear trendlines shown in Fig.~\ref{fig:line-widths}. The Pearson correlation coefficient $r$ has been computed for each pair of populations. The dashed black line represents the one-to-one correlation line.}
\label{fig:width-correlations}
\end{figure*}

By eye, there are also hints that the relative width of the four different objects in the same MKID pixel --- continuous inductor, non-continuous inductor, long capacitor, and short capacitor --- are correlated beyond the radial trend for that population. To examine this in more detail, we first subtract the trendline value from each population. The trendline-subtracted line widths are plotted against one another for each relevant pair of objects in Fig.~\ref{fig:width-correlations}a (for chip one) and Fig.~\ref{fig:width-correlations}b (for chip two). With the data in this form it is clear that the objects on a given MKID pixel vary above or below the linear trend together. This evidence is further supported by the large, positive Pearson correlation coefficients $r$ for each pairing of objects (displayed in the figure's legend). This information is fed back into the fabrication process to help improve the quality of future fabrications.

%\begin{figure*}
%\centering
%\includegraphics[width=1.0\columnwidth]{figures/detectors/width_trends.pdf}
%\caption{Line width correlation between the objects within a single MKID pixel, relative to their linear trendlines shown in Fig.~\ref{fig:line-widths}.}
%\label{fig:width-trends}
%\end{figure*}

%\begin{figure*}
%\centering
%\includegraphics[width=0.8\columnwidth]{figures/detectors/width_correlations_chip_two.pdf}
%\caption{A visualization of the line width correlation between the objects within a single MKID pixel on chip two, relative to their linear trendlines shown in Fig.~\ref{fig:line-widths-two}. The Pearson correlation coefficient $r$ has been computed for each pair of populations.}
%\label{fig:width-correlations-two}
%end{figure*}

\section{Conclusions}
We have developed a pipeline for imaging cryogenic superconducting detectors and analyzing said images to search for visual defects, measure conducting line width, and give anticipated performance data that can be used to detect and discard problematic wafers before the time and expense of cryogenic testing. Using a suite of simulations, we have demonstrated that the pipeline can find common kinds of defects with an accuracy of $98.6\%$; in testing on actual detector micrographs we have also found that it effectively identifies unanticipated problems (deposited material) as well.

We have applied the pipeline to analyze images of prototype MKIDs for the planned SPT-$3$G$+$ experiment, finding a rich population of defects and using information about said defects to estimate the operability of each detector. We predict a yield equal to $100\%$ for chip one and between $38.3\%$ and $45.1\%$ for chip two, depending on how one chooses to interpret the overflowing detectors. Although these predictions are not, and are not intended to be, in perfect agreement with cryogenic measurements of each chip's yield, they provide valuable information that would, in this case, have recommended the well-performing chip one for cryogenic testing and identified an unanticipated fabrication problem (material deposition) for chip two that would otherwise not have been caught even in first-round cryogenic testing. Furthermore, our measurements of detector conducting line width have provided evidence of a directionality bias in the MKID fabrication process that was unknown previous to this work; the techniques here have thus already had an effect on our detector fabrication process even on only two prototype chips from the research and development phase. While we have tailored the pipeline development to this specific hardware, the process is general and can be applied to other geometries and types of detectors, both within and outside of the CMB field.

\subsection{Future work}
There are a number of improvements that could further advance this pipeline. The pipeline is not currently configured to detect addition-type defects in the MKID inductors. Adding this functionality would improve the ability to predict shifts in the resonant frequency of the detector. Expanding the defect search to the lines connecting the inductors and capacitors within an MKID pixel would increase the accuracy of our yield predictions, while expanding the search to the feedlines running through the full chip would be a major improvement, as currently the entire feedline is inspected by eye. While this latter addition would present a number of computing and analysis difficulties, it would significantly reduce the time spent on manual inspection. Lastly, a more thorough comparison to the cryogenic characterization data (similar to what was done in our previous study utilizing TES bolometers~\citep{ferguson22b}) could prove illuminating, potentially revealing new avenues for further study.

Beyond the imaging and analysis pipeline itself, efforts are ongoing to improve the connection between observed features and performance in the prototype MKID detectors. A detailed modeling framework for resonant frequencies based on device geometry is in development. Additional characterization of the MKIDs is also underway to provide discrete mapping between each MKID and its resonance in the measured transmission curve. Connecting observed visual features, such as relative line width, together with the model predictions and additional measurements will greatly improve the ability to inform future detector design and fabrication as these devices continue to develop.

\section*{Acknowledgments}
%\markboth{ACKNOWLEDGMENTS}{ACKNOWLEDGMENTS}
%The authors would like to thank Claudio Kopper, Matthew Becker, Nesar Ramachandra, and Markus Rau for their helpful conversations, as well as 
The authors would like to thank the SPT-$3$G$+$ collaboration for the use of their detector chips for this study. The South Pole Telescope program is supported by the National Science Foundation (NSF) through grants PLR-1248097, OPP-1852617. Argonne National Laboratory's work was supported by the U.S. Department of Energy, Office of High Energy Physics, under contract DE-AC02-06CH11357. Work performed at the Center for Nanoscale Materials, a U.S. Department of Energy Office of Science User Facility, was supported by the U.S. DOE, Office of Basic Energy Sciences, under contract DE-AC02-06CH11357. We gratefully acknowledge the computing resources provided on Crossover, a high-performance computing cluster operated by the Laboratory Computing Resource Center at Argonne National Laboratory. This work made use of the Pritzker Nanofabrication Facility, part of the Pritzker School of Molecular Engineering at the University of Chicago, which receives support from Soft and Hybrid Nanotechnology Experimental (SHyNE) Resource (NSF ECCS-2025633), a node of the National Science Foundation's National Nanotechnology Coordinated Infrastructure [RRID: SCR\_022955]. KRF and KRD acknowledge support from the U.S. Department of Energy's Office of Science Graduate Student Research (SCGSR) Program. PSB acknowledges support through the UKRI Future Leaders Fellowship Program (MR/W006499/1). This work makes use of the \texttt{numpy}~\citep{harris20}, \texttt{matplotlib}~\citep{hunter07}, \texttt{scipy}~\citep{virtanen20}, \texttt{scikit-image}~\citep{vanderwalt14}, and \texttt{scikit-learn}~\citep{pedregosa11} Python packages.

% References
%\def\newblock{\hskip .11em plus .33em minus .07em}
\bibliographystyle{prd} % makes bibtex use prd.bst
\bibliography{refs}

\end{document}